\documentclass[prb,twocolumn,preprintnumbers,amsmath,amssymb,superscriptaddress]{revtex4}
\usepackage{bm}

\usepackage{graphicx}

\usepackage{color}
\usepackage{bm}
\usepackage{epstopdf}

\newcommand \be{\begin{eqnarray}}
\newcommand \ee{\end{eqnarray}}
\newcommand \ba{\begin{align}}
\newcommand \eea{\end{align}}

\newcommand {\p}[1]{\partial_{#1}}

\newcommand \bd{\mathbf}

\begin{document}

\title{Spin multistability in dissipative polariton channels}

\author{\"{O}. Bozat}
\altaffiliation{Current Address: Faculty of Engineering and Natural Sciences, Sabanc{\i} University, Orhanl{\i} - Tuzla, 34956, Turkey}
\affiliation{International Institute of Physics, Av. Odilon
Gomes de Lima, 1772, Capim Macio, 59078-400, Natal, Brazil}

\author{I. G. Savenko}
\affiliation{Science Institute, University of Iceland, Dunhagi 3,
IS-107, Reykjavik, Iceland}
\affiliation{Division of Physics and
Applied Physics, Nanyang Technological University 637371, Singapore}

\author{I. A. Shelykh}
\affiliation{Science Institute, University of Iceland, Dunhagi 3,
IS-107, Reykjavik, Iceland}
\affiliation{Division of Physics and Applied Physics, Nanyang Technological University
637371, Singapore}
\date{\today}

\begin{abstract}
We present a model for theoretical description of the dynamics of
a system of spinor cavity polaritons in real space and time
accounting for all relevant types of the interactions and effective
magnetic fields. We apply our general
formalism for the consideration of the polarization dynamics of the
coherently driven one dimensional polariton channel. We investigate
effect of the temperature and the longitudinal-transverse splitting
on the spin (polarization) multistability and hysteresis arising
from the polarization-dependent polariton-polariton interaction. We
show that the effect of the phase of the driving laser pump is
as important as its strength and demonstrate that the
multistability behavior can survive up to high temperatures in
presence of longitudinal-transverse splitting.
\end{abstract}

\maketitle


\section{Introduction}

Cavity polaritons are composite particles, arising from strong
coupling between photonic mode of a planar semiconductor microcavity
and exciton transition in Quantum Well (QW), embedded in the cavity
at the point where the electric field of the confined
electromagnetic state reaches its maximum. Having a hybrid
half-light - half-matter nature cavity polaritons demonstrate a set
of peculiar properties which make them different from other
quasiparticles in solid state systems. Extremely low effective mass
of the cavity polaritons (about $10^{-4}-10^{-5}$ of the free
electron mass) together with strong polariton-polariton
interactions makes polaritonic system an ideal candidate for
observation of the variety of quantum collective phenomena at
surprisingly high temperatures. Achievement of polariton Bose-Einstein 
condensation was first reported at $T=20 K$
\cite{KasprzakNature} and later on even at room temperature
\cite{Christopoulos}. Later on polariton superfluidity
\cite{AmoNature}, Josephson effect \cite{LagoudakisJosephson} and
formation of topological excitations \cite{Vortex} were
experimentally observed. Other theoretically predicted effects such
as polariton self-trapping \cite{ShelykhJosephson},
polariton-mediated superconductivity \cite{LaussySupercond} still
wait for their experimental confirmation.

\begin{figure}
\includegraphics[width=0.95\linewidth]{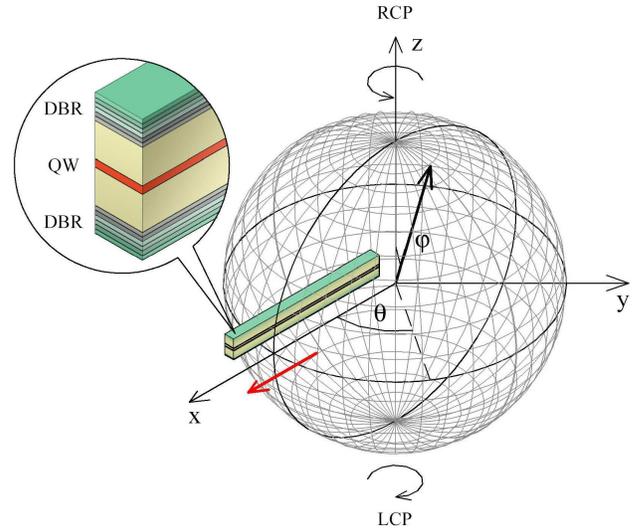}
\caption{Sketch of the system showing the position of the polariton
channel with respect to the Poincar\'{e} sphere (also known as Bloch
sphere and pseudospin sphere). The latter serves to illustrate
different possible polarizations of the light. If the vector of the
pseudospin lies in the xy plane, the light is linearly polarized and
if it is parallel to z-axes it is circularly polarized. Other
orientations correspond to the general case of elliptical
polarization. Red arrow shows the direction of the effective
magnetic field created by the longitudinal-transverse splitting
$\Omega$.} \label{fig1}
\end{figure}

Besides fundamental interest, quantum microcavities in the strong
coupling regime can be used for optoelectronic applications
\cite{ReviewPolaritonDevices}. For more then a decade, the only
object of the study in this context was polariton laser
\cite{Imamoglu}- a novel type of the coherent emitter which explores
the possibility of the polariton BEC. In the last years, however,
the emphasis became to shift on other types of the devices based on
transport properties of cavity polaritons in real space. It was
noticed that the peculiar spin structure of polaritons opens a way
for creation of optical analogs of spintronic components (so-called
spinoptronic devices \cite{ShelykhSpinoptronics}). With respect to
optics, spin-optronics has the advantage of being able to use
particle-particle interactions occurring in nanostructures and
resulting in strong nonlinearities. With respect to spintronics, it
has the advantage of strongly reducing the dramatic impact of
carrier spin relaxation or decoherence, which has severely limited
the achievement or the functionality of any working
semiconductor-based spintronic devices.

In this context, the analysis of one-dimensional (1D) polariton
transport is of particular importance \cite{WertzNature}, as 1D
polariton channels are fundamental building blocks of such future
spinoptronic devices as polariton neurons \cite{LiewNeuron} and
polariton integrated circuits \cite{LiewCircuit}. It should be
noted, that current state of growth technology offers a large
variety of methods of the lateral confinement of cavity polaritons
\cite{Confinement} and polariton quantum wires (1D polariton
channels) can be routinely produced.

Currently, the theoretical study of transport of spinor cavity
polaritons in the real space is based on the assumption of full
coherence of the polaritonic system. Polariton-polariton
interactions are either neglected
\cite{SpinHall,Glazov,ShelykhBerry}, either treated within
frameworks of spinor Gross-Pitaevskii equations (GPe) \cite{ShelykhGP}.
The non-coherent processes coming from the interaction of the
polaritonic system with a phonon bath in most cases are not
accounted for or treated within simple phenomenological models
lacking microscopic justification
\cite{Wouters2007,BerloffVortex,Tim1D}. On the other hand, it is
clear that polariton-phonon interaction is of crucial importance,
as it provides a thermalization mechanism for a polaritonic ensemble
and drastically affects such experimentally observable quantities as
first and second order coherencies \cite{Sarchi,Laussy2004}.

It should be noted, that for spatially homogeneous polariton system
polariton-phonon interactions can be accounted for using a system of
the semiclassical Boltzmann equations
\cite{Porras2002,Kasprzak2008,Haug2005,Cao}. This method, however,
has several serious drawbacks. First, it is based on the assumption
that the system is fully incoherent, and thus variety of
intriguing nonlinear phenomena such as bistability and
multistability can not be described. Second, it provides the
information about occupation numbers in the reciprocal space only,
and thus can not be used for description of the dynamics of the
spatially inhomogeneous system. This makes this formalism
inappropriate for modeling of spinoptronic devices based on
polariton transport in real space.

In the present paper we present formalism suitable for the
description of the dynamics of an inhomogeneous spinor polariton
system in real space and time accounting for all relevant types of
the processes. Namely, we take into account polariton-polariton and
polariton-phonon interactions and effective longitudinal-transverse
(TE-TM) magnetic field acting on polariton spin.  Our consideration
is based on the Lindblad approach for density matrix dynamics and
represents a generalization of our previous work where spinless case
was considered \cite{Savenko,Magnusson2011}. We use our results for
modeling of the spin dynamics of the polaritons in 1D channels,
investigating the role played by decoherence at different
temperatures.


\section{Formalism}

We describe the state of the system (polaritons
plus phonons) by its density matrix $\chi$, for which we apply the
Born approximation factorizing it into the phonon part which is supposed
to be time-independent and corresponds to the thermal distribution
of acoustic phonons
$\chi_{ph}=\texttt{exp}\left\{-\frac{{\cal H}_{ph}}{k_BT}\right\}$ and
the polariton part $\chi_{pol}$ whose time dependence should be
determined,  $\chi=\chi_{ph}\otimes \chi_{pol}$.  Our aim is to find
 dynamic equations for the time evolution of the single-particle
polariton density matrix in real space and time
\begin{eqnarray}
\rho_{\sigma,\sigma'}(\textbf{r},\textbf{r}',t)&=&Tr\left\{\widehat{\psi}^\dagger_\sigma(\textbf{r},t)\widehat{\psi}_{\sigma'}(\textbf{r}',t)\rho\right\}\\
\nonumber
&=&\langle\widehat{\psi}^\dagger_\sigma(\textbf{r},t)\widehat{\psi}_{\sigma'}(\textbf{r}',t)\rangle
\end{eqnarray}
where
$\widehat{\psi}^\dagger_{\sigma}(\textbf{r},t),\widehat{\psi}_{\sigma'}(\textbf{r},t)$
are operators of the spinor polariton field, the subscripts
$\sigma,\sigma'=\pm1$ denote the z-projection of the spin of cavity
polaritons and correspond to right-and left-circular polarized
states and the trace is performed by all the degrees of freedom of
the system. The particularly interesting quantities are matrix
elements with $\textbf{r}=\textbf{r}'$ which give the density and
polarization of the polariton field in real space and time
\begin{eqnarray}
&&n(\textbf{r},t)=\sum\limits_{\sigma=\pm1}\rho_{\sigma,\sigma}(\textbf{r},\textbf{r},t),\\
&&s_z(\textbf{r},t)=\frac{1}{2}\left[\rho_{+1,+1}(\textbf{r},\textbf{r},t)-\rho_{-1,-1}(\textbf{r},\textbf{r},t)\right],\\
&&s_x(\textbf{r},t)+is_y(\textbf{r},t)=\rho_{+1,-1}(\textbf{r},\textbf{r},t).
\end{eqnarray}
The off-diagonal matrix elements with $\textbf{r}\neq\textbf{r}'$ also have physical meaning and describe spatial coherence in the system.

To obtain expressions for temporal dynamics of the components of
single particle density matrix it is convenient to go to the
reciprocal space, making a Fourier transform of the one-particle
density matrix,
\begin{align}
\nonumber
\rho_{\sigma,\sigma'}(\textbf{k},\textbf{k}',t)&=(2\pi)^d/L^d\int e^{i(\textbf{kr}-\textbf{k}'\textbf{r}')}\rho_{\sigma,\sigma'}(\textbf{r},\textbf{r}',t)d\textbf{r}d\textbf{r}'\\
&=Tr\left\{a_{\sigma,\textbf{k}}^+a_{\sigma',\textbf{k}'}\chi\right\}\equiv\langle a_{\sigma,\textbf{k}}^+a_{\sigma',\textbf{k}'}\rangle,
\end{align}
where $d$ is the dimensionality of the system ($d=2$ for
non-confined polaritons, $d=1$ for the polariton channel), L is its
linear size,  $a_{\sigma \textbf{k}}^+$, $a_{\sigma \textbf{k}}$ are
creation and annihilation operators of the polaritons with circular
polarization $\sigma$ and momentum \textbf{k}. Note, that we have
chosen the prefactor in a Fourier transform in such a way, that the
values of $\rho(\textbf{k},\textbf{k}',t)$ are dimensionless, and
diagonal matrix elements give occupation numbers of the states in
discretized reciprocal space. Knowing the density matrix in
reciprocal space, we can find the density matrix in real space
straightforwardly applying the inverse Fourier transform.

The total Hamiltonian of the system can be represented as a sum of two parts,
\begin{equation}
{\cal H}={\cal H}_1+{\cal H}_2,
\end{equation}
where the first term ${\cal H}_1$ describes the "coherent" part of the
evolution, corresponding to free polariton propagation,
polariton-polariton interactions and the effect of TE-TM splitting,
and the second term ${\cal H}_2$ corresponds to the dissipative interaction
with acoustic phonons. The two terms affect the polariton density
matrix in a qualitatively different way.


\subsection{Polariton-polariton interactions}

The part of the evolution
corresponding to ${\cal H}_1$ is given by the following expression
\begin{eqnarray}
{\cal H}_1&=&\sum_{\bd k\sigma}E_{\bd k}a_{\bd k\sigma}^+a_{\bd k,\sigma}+\sum_{\bd k,\sigma} \Omega(\textbf{k})a_{\bd k,\sigma}^+a_{\bd k,-\sigma}\\
\nonumber
&+&U_{1}\sum_{\bd {k_1},\bd{k_2},\bd{p},\sigma} a_{\bd{k_1},\sigma}^+ a_{\bd{k_2},\sigma}^+ a_{\bd{k_1}+\bd p,\sigma}a_{\bd{k_2}-\bd{p},\sigma}\\
\nonumber
&+&U_{2}\sum_{\bd{k_1},\bd{k_2},\bd{p},\sigma} a_{\bd{k_1},\sigma}^+ a_{\bd{k_2},-\sigma}^+ a_{\bd{k_1}+\bd{p},\sigma}a_{\bd{k_2}-\bd{p},-\sigma},
\end{eqnarray}
where $E_\textbf k$ gives the dispersion of the polaritons,
$\Omega(\textbf{k})$ is the TE-TM splitting corresponding to the in-
plane effective magnetic field leading to the rotation of the
pseudospin of cavity polaritons, $U_1$ is the matrix element of the
interaction between polaritons of the same circular polarization,
$U_2$-matrix element of the interaction between polaritons of
opposite circular polarizations. In the current paper we neglect the
\textbf{p}-dependence of the polariton-polariton interaction
constant coming from Hopfield coefficients for simplicity. As well,
we will suppose  $\Omega(\textbf{k})=const$ which corresponds well
to the situation of the polariton channel (but not for 2D polariton
system).

The effect of ${\cal H}_1$ on the evolution of the density matrix is described by the
Liouville-von Neumann equation,
\begin{equation}
i\hbar\left(\partial_t\chi\right)^{(1)}=\left[{\cal H}_1;\chi\right],
\label{liouville}
\end{equation}
which after the use of the mean field approximation leads to the
following dynamic equations for the elements of the single-particle
density matrix in the reciprocal space (the derivation is completely
analogical to those presented in Ref.~\onlinecite{Savenko}):
\begin{widetext}
\begin{align}
-i\hbar \{\p t \rho_{\sigma,\sigma'}(\bd k,\bd k')\}^{(1)} &=( E_\bd k-E_{\bd k'})\rho_{\sigma,\sigma'}(\bd k,\bd k') +\Omega\left[ \rho_{-\sigma,\sigma'} (\bd k,\bd k')-\rho_{\sigma,-\sigma'} (\bd k,\bd k') \right]\\
\nonumber
&+U_{1}\sum_{\bd {k_1},\bd p}\left[ \rho_{\sigma,\sigma}(\bd {k_1},\bd {k_1}-\bd p)\rho _{\sigma,\sigma'}(\bd k-\bd p,\bd k',)- \rho_{\sigma',\sigma'}(\bd {k_1},\bd {k_1}-\bd p) \rho_{\sigma,\sigma'}(\bd k,\bd k'+\bd p)\right] \\
\nonumber
&+U_{2}\sum_{\bd {k_1},\bd p}\left[ \rho_{-\sigma,-\sigma}(\bd {k_1},\bd {k_1}-\bd p)\rho_{\sigma,\sigma'} (\bd k-\bd p,\bd k')- \rho_{-\sigma',-\sigma'}(\bd {k_1},\bd {k_1}-\bd p) \rho_{\sigma,\sigma'}(\bd k,\bd k'+\bd p)\right].
\end{align}
\end{widetext}


\subsection{Scattering with acoustic phonons}

Polariton-phonon scattering corresponds to the interaction of the
quantum polariton system with the classical phonon reservoir. It is
of dissipative nature, and thus straightforward application of the
Liouville-von Neumann equation is impossible. One should rather use
the approach based on the Lindblad formalism, which is standard in
quantum optics and results in the master equation for the full
density matrix of the system \cite{Carmichael}. For the convenience
of the reader, we give the main steps of the derivation of the
dissipative part of dynamic equations for spinor polariton system,
omitting however all technical details which can be found elsewhere
\cite{Savenko}.

The Hamiltonian of the interaction of polaritons with acoustic
phonons in Dirac picture can be represented as
\begin{align}
&{\cal H}_2(t)={\cal H}^-(t)+{\cal H}^+(t)=\\
\nonumber&=\sum_{\sigma,\textbf{k},\textbf{q}}D(\textbf{q})e^{i(E_{\textbf{k}+\textbf{q}}-E_{\textbf{k}})t}a_{\sigma,\textbf{k}+\textbf{q}}^+a_{\sigma,\textbf{k}}(b_\textbf{q}e^{-i\omega_{\textbf{q}}t}+b_{-\textbf{q}}^+e^{i\omega_{\textbf{q}}t}),
\end{align}
where $a_{\sigma\textbf{k}}$ are operators for spinor polaritons,
$b_\textbf{q}$ operators for spinless phonons, $E_\textbf{k}$ and
$\omega_\textbf{q}$ are dispersion relations for polaritons and
acoustic phonons respectively, $D(\textbf{q})$ is the
polariton-phonon coupling constant. In the last equality we
separated the terms ${\cal H}^+$ where a phonon is created, containing the
operators $b^+$, from the terms ${\cal H}^-$ in which it is destroyed,
containing operators $b$.

Now, one can consider a hypothetical situation when
polariton-polariton interactions are absent, and the redistribution
of the polaritons in reciprocal space is due to the scattering with
acoustic phonons only. One can rewrite the Liouville-von Neumann
equation in an integro-differential form and apply the so called
Markovian approximation, corresponding to the situation of fast
phase memory loss (see Ref.~\onlinecite{Carmichael} for the details
and discussion of limits of validity of the approximation)
\begin{align}
\label{Liouville_int}
\left(\partial_t\chi\right)^{(2)}&=-\frac{1}{\hbar^2}\int_{-\infty}^t dt'\left[{\cal H}_{2}(t);
  \left[{\cal H}_{2}(t');\chi(t)\right]\right]\\
\nonumber &= \delta_{\Delta E}\left[2\left({\cal H}^+\chi {\cal H}^-+{\cal H}^-\chi\right.
{\cal H}^+\right)   \nonumber \\
&\left.-\left({\cal H}^+{\cal H}^-+{\cal H}^-{\cal H}^+\right)\chi-\chi\left({\cal H}^+{\cal H}^-+{\cal H}^-{\cal H}^+\right)\right], \nonumber
\end{align}
where the coefficient $\delta_{\Delta E}$ denotes energy
conservation and has dimensionality of inverse energy and in the
calculation taken to be equal to the broadening of the polariton
state \cite{KavokinMalpuech}. For time evolution of the mean value
of any arbitrary operator $ \langle \widehat{A}\rangle=Tr(\chi
\widehat{A})$ due to scattering with phonons one thus has
(derivation of this formula is represented in
Ref.~\onlinecite{Savenko}):
\begin{equation}\label{eqM}
\left\{\partial_t\langle
\widehat{A}\rangle\right\}^{(2)}=\delta_{\Delta E}\left(\langle[{\cal H}^-;[\widehat{A};{\cal H}^+]]\rangle+\langle[{\cal H}^+;[\widehat{A};{\cal H}^-]]\rangle\right).
\end{equation}
Putting $\widehat{A}=a_{\sigma,\textbf{k}}^+a_{\sigma',\textbf{k}'}$  in this equation
we get the contributions to the dynamic equations for the elements
of the single-particle density matrix coming from polariton-phonon
interaction:
\begin{widetext}
\begin{align}
\{\p t n_{\bd k,\sigma}\}^{(2)}&\\
\nonumber
=&\sum_{\bd q, E_\bd k<E_{\bd k+\bd q} } 2 W(\bd q) \left[ (n_{\bd k,\sigma }+1) n_{\bd k+\bd q,\sigma }(n_\bd q ^{ph}+1)-n_{\bd k,\sigma }( n_{\bd k+\bd q,\sigma }+1)n_\bd q ^{ph} \right. \\
\nonumber
+&\left. \frac 1 2 (\rho_{\sigma,-\sigma}(\bd k,\bd k)\rho_{-\sigma,\sigma}(\bd k+\bd q,\bd k+\bd q)+\rho_{-\sigma,\sigma}(\bd k,\bd k)\rho_{\sigma,-\sigma}(\bd k+\bd q,\bd k+\bd q)) \right] \\
\nonumber
+&\sum_{\bd q, E_\bd k>E_{\bd k+\bd q} } 2 W(\bd q) \left[ (n_{\bd k,\sigma }+1) n_{\bd k+\bd q,\sigma }n_{-\bd q} ^{ph}-n_{\bd k,\sigma }( n_{\bd k+\bd q,\sigma }+1)(n_{-\bd q} ^{ph}+1 ) \right. \\
\nonumber
-&\left. \frac 1 2 (\rho_{\sigma,-\sigma}(\bd k,\bd k) \rho_{-\sigma,\sigma}(\bd k+\bd q,\bd k+\bd q)+\rho_{-\sigma,\sigma}(\bd k,\bd k)\rho_{\sigma,-\sigma} (\bd k+\bd q,\bd k+\bd q)) \right] ,
\end{align}
\begin{align}
& \{\p t \rho_{\sigma,-\sigma}(\bd k,\bd k)\}^{(2)}\\
\nonumber
=&\sum_{\bd q, E_\bd k<E_{\bd k+\bd q} } 2 W(\bd q) \left[ \rho _{\sigma,-\sigma}(\bd k,\bd k)  \left( \frac 1 2 (n_{\bd k +\bd q, \sigma}+n_{\bd k +\bd q, \sigma'})-n_\bd q ^{ph} \right)+ \rho _{\sigma,-\sigma}(\bd k+\bd q,\bd k+\bd q) \left(\frac 1 2 (n_{\bd k,\sigma}+n_{\bd k,\sigma'})+n_\bd q ^{ph} +1 \right) \right] \\
\nonumber
-&\sum_{\bd q, E_\bd k>E_{\bd k+\bd q} } 2 W(\bd q) \left[ \rho_{\sigma,-\sigma} (\bd k,\bd k) \left( \frac 1 2 (n_{\bd k +\bd q, \sigma}+n_{\bd k +\bd q, \sigma'})+n_{-\bd q} ^{ph}+1 \right) + \rho_{\sigma,-\sigma} (\bd k+\bd q, \bd k+\bd q) \left(\frac 1 2 (n_{\bd k,\sigma}+n_{\bd k,\sigma'})-n_{-\bd q} ^{ph} \right) \right],
\end{align}
\begin{align}
 \{\p t \rho_{\sigma,\sigma}(\bd k,\bd k')\}^{(2)}&=\sum_{\bd q, E_\bd k<E_{\bd k+\bd q} }  W(\bd q) \left[
\rho_{\sigma,\sigma}(\bd k,\bd k') (n_{\bd k + \bd q,\sigma}-n_\bd q ^{ph}) +\rho_{-\sigma,\sigma}(\bd k,\bd k') \rho_{\sigma,-\sigma}(\bd k+\bd q,\bd k+\bd q)\right] \\ \nonumber
&-\sum_{\bd q, E_\bd k>E_{\bd k+\bd q} }  W(\bd q) \left[
\rho_{\sigma,\sigma}(\bd k,\bd k')
 (n_{\bd k + \bd q,\sigma}+n_\bd q ^{ph}+1) +\rho_{-\sigma,\sigma}(\bd k,\bd k')  \rho_{\sigma,-\sigma}(\bd k+\bd q,\bd k+\bd q) \right] \\ \nonumber &+\sum_{\bd q, E_{\bd k'}<E_{\bd k'+\bd q} }  W(\bd q) \left[
\rho_{\sigma,\sigma}(\bd k,\bd k') (n_{\bd k' + \bd q,\sigma}-n_\bd q ^{ph}) +\rho_{\sigma,-\sigma}(\bd k,\bd k') \rho_{-\sigma,\sigma}(\bd k'+\bd q,\bd k'+\bd q)\right] \\ \nonumber
&-\sum_{\bd q, E_{\bd k'}>E_{\bd k'+\bd q} }  W(\bd q) \left[
\rho_{\sigma,\sigma}(\bd k,\bd k')
 (n_{\bd k' + \bd q,\sigma}+n_\bd q ^{ph}+1) +\rho_{\sigma,-\sigma}(\bd k,\bd k')  \rho_{-\sigma,\sigma}(\bd k'+\bd q,\bd k'+\bd q) \right],
\end{align}

\begin{align}
\{\p t \rho_{\sigma,-\sigma}(\bd k,\bd k')\}^{(2)}&=\sum_{\bd q, E_\bd k<E_{\bd k+\bd q} }  W(\bd q) \left[
\rho_{\sigma,-\sigma}(\bd k,\bd k' ) (n_{\bd k + \bd q,\sigma}-n_\bd q ^{ph}) +\rho_{-\sigma,-\sigma}(\bd k,;\bd k') \rho_{\sigma,-\sigma}(\bd k+\bd q,\bd k+\bd q)\right] \\ \nonumber
&-\sum_{\bd q, E_\bd k>E_{\bd k+\bd q} }  W(\bd q) \left[
\rho_{\sigma,-\sigma}(\bd k,\bd k')
 (n_{\bd k + \bd q,\sigma}+n_\bd q ^{ph}+1) +\rho_{-\sigma,-\sigma}(\bd k,\bd k')  \rho_{\sigma,-\sigma}(\bd k+\bd q,\bd k+\bd q) \right] \\ \nonumber &+\sum_{\bd q, E_{\bd k'}<E_{\bd k'+\bd q} }  W(\bd q) \left[
\rho_{\sigma,-\sigma}(\bd k,\bd k') (n_{\bd k' + \bd q,\sigma'}-n_\bd q ^{ph}) +\rho_{\sigma,\sigma}(\bd k,\bd k') \rho_{\sigma,-\sigma}(\bd k'+\bd q,\bd k'+\bd q)\right] \\ \nonumber
&-\sum_{\bd q, E_{\bd k'}>E_{\bd k'+\bd q} }  W(\bd q) \left[
\rho_{\sigma,-\sigma}(\bd k,\bd k')
 (n_{\bd k' + \bd q,\sigma'}+n_\bd q ^{ph}+1) +\rho_{\sigma,\sigma}(\bd k,\bd k')  \rho_{\sigma,-\sigma}(\bd k'+\bd q,\bd k'+\bd q) \right],
\end{align}
\end{widetext}
where, $n_{\bd k, \sigma}=\rho_{\sigma,\sigma}(\bd k,\bd k)$,
$\rho_{+1;-1}(\bd k)=s_x(\bd k)+is_y(\bd k)$ and $W(\textbf{q})$
denote spin-independent scattering rates with acoustic phonons (see
Ref.~\onlinecite{Savenko} for the details). The first two equations
corresponding to $\textbf{k}=\textbf{k}'$ are nothing but the spinor
Boltzmann equations for polariton-phonon scattering describing the
redistribution of the polaritons in the reciprocal space which were
obtained earlier using another techniques \cite{KKavokinBoltzmann}.
The equations for off-diagonal matrix elements with
$\textbf{k}\neq\textbf{k}'$ describe their decay which physically
corresponds to the decay of the coherence in the system coming from
polariton-phonon interactions.


\subsection{Pumping terms}

In this paper we concentrate on the case when system is pumped by
external coherent laser beam. The corresponding Hamiltonian can be
introduced as
\begin{equation}
{\cal H}_{cp}=\sum_{\bd k,\sigma}p_{\bd k, \sigma}(t) a_{\bd k, \sigma}^+
+h.c.
\end{equation}
Here $p_\bd k$ is the Fourier transform of the pumping amplitude in real space
\begin{equation}
p_\sigma(\bd x,t)=P_\sigma(\bd x)e^{i\bd k_p \bd x}e^{-i\omega_p t},
\end{equation}
where $P_\sigma(\bd x)$ is the pumping spot profile in real space,
$\bd k_p$ is an in-plane pumping vector resulting from the
inclination of the laser beam as respect to the vertical, and
$\omega_p$ is the pumping frequency of the single-mode laser. Time
evolution of the arbitrary element of density matrix is given by \cite{Magnusson2011}
\begin{equation}
\{\p t \rho_{\sigma,\sigma'}(\bd k,\bd k')\}^{(cp)}=\frac i \hbar
(p_{\bd k, \sigma} ^\ast(t) \langle a_{\bd k',\sigma '} \rangle
-p_{\bd k', \sigma '}(t) \langle a_{\bd k,\sigma} \rangle ^* ),
\end{equation}
where the time evolution of the mean values of the annihilation
operator reads
\begin{widetext}
\begin{align}
\p t \langle a_{\bd k, \sigma}\rangle &= -\frac i \hbar p_{\bd k,
\sigma}(t)-\frac i \hbar E_\bd k \langle a_{\bd k,\sigma} \rangle
-\frac i \hbar \Omega_\bd k \langle a_{\bd k, -\sigma}\rangle \\
\nonumber &-\frac i \hbar \sum_{\bd k_2,p} (
U_1\rho_{\sigma,\sigma}(\bd k_2,\bd k_2-\bd p)  + U_2
\rho_{-\sigma,-\sigma}(\bd k_2,\bd k_2-\bd p)  )\langle a_{\bd k
+\bd p, \sigma} \rangle \\ \nonumber & +\sum_{\bd q, E_\bd k <
E_{\bd k + \bd q} } W(\bd q) \left[(n_{\bd k + \bd q,\sigma}-n_\bd
q^{ph})\langle a_{\bd k,\sigma}\rangle +\rho_{-\sigma,\sigma}(\bd
k+\bd q,\bd k+\bd q)\langle a_{\bd k, -\sigma}\rangle\right] \\
\nonumber & - \sum_{\bd q, E_\bd k > E_{\bd k + \bd q} }   W(\bd q)
\left[(n_{\bd k + \bd q,\sigma}+n_\bd q^{ph}+1)\langle a_{\bd
k,\sigma}\rangle +\rho_{-\sigma,\sigma}(\bd k+\bd q,\bd k+\bd
q)\langle a_{\bd k, -\sigma}\rangle\right].
\end{align}
\end{widetext}
%


\subsection{Dynamics of the polarization}

The dynamics of the circular polarization degree $\wp_{c}$ of the light emission from the ground $k=0$ state can be defined as
\begin{equation}
\wp_{c}=\frac{n^+_{\textbf{k}=0}-n^-_{\textbf{k}=0}}{n^+_{\textbf{k}=0}+n^-_{\textbf{k}=0}},
\end{equation}
where $n^+_{\textbf{k}=0}(t)$ and $n^-_{\textbf{k}=0}(t)$ stand for
the populations of polaritons with pseudospin $\pm 1$
correspondingly in the ground state of the dispersion.

One should mention the effect of longitudinal-transverse splitting
$\mathbf\Omega$ on the polarization degree dynamics, since it couples
$\sigma^+$ and $\sigma^-$ modes together. Its role becomes more
evident if one switches to the pseudospin formalism which is
described in the introductory part of the manuscript. From the
formal point of view, TE-TM splitting is equivalent to the effective
magnetic field in +x direction $\bd \Omega=\bd e_x\Omega$ (along the
quantum wire). In the same time, the polariton-polariton interaction
gives rise to another effective magnetic field oriented in z
direction (structure growth axes) $\bd \Omega_{p-p} \propto \bd e_z
(U_1-U_2)(n^+ - n^-)$ (see Ref.~\onlinecite{ReviewSpin}). Therefore, the
total effective magnetic field represents superposition $\bd
\Omega_{tot}=\bd \Omega_{p-p}+\bd \Omega$. Accordingly, it is
possible to rewrite the kinetic equations as a coupled equations for
occupation number $n_{\sigma}$ and in-plane pseudospin $\bd S_{\bot
}$ \cite{Shelykh05}. Considering only the effect of effective
magnetic fields (assuming infinite lifetime and absence of interaction with phonons), coupled equations are given as
\begin{align}\label{pseudo1}
\partial_t n^+_{\textbf{k}=0} &\propto \bd e_z \cdot (\bd S_\bot \times \bd \Omega) , \\
\partial_t \bd S_\bot &\propto (\bd S_\bot \times \bd \Omega_{p-p})+\frac{1}{2}(n^+_{\textbf{k}=0} - n^-_{\textbf{k}=0}) \bd \Omega.\label{pseudo2}
\end{align}
This corresponds to the precession of the pseudospin along time
dependent magnetic field, which leads to its non-trivial dynamics.
%

\section{Results and discussion}

We consider a microcavity based on AlGaAs family of alloys and use
the following parameters. The Rabi splitting was taken equal to 15
meV, polariton effective mass 3$\times $10$^{-4}$ of the free
electron mass and  detuning between the pure photonic and excitonic
modes 3 meV. The polaritonic quantum wire is 50 $\mu$m long and 2
$\mu$m wide. Further, we use typical polariton lifetime in a medium
Q-factor microcavity, $\tau=2$ ps. The polariton-polariton and
polariton-phonon scattering rates have been taken independent on the
wavevector for simplicity. The matrix element of polariton-polariton
interaction was estimated using expression $U\approx 3E_ba_B^2/S$,
where $E_b$ is the exciton binding energy, $a_B$ is its Bohr radius
and $S$ is the area of the wire, which gives $U\approx20$ neV. The
polariton-phonon scattering rate $W=10^{8}$ s$^{-1}$. Pump laser is
detuned above the energy of the lower polariton branch by $\delta=1$
meV. We consider the case of spatially homogeneous cw pump of
different polarizations.

The bistable (for spinless condensate) and multistable (if one
accounts for the spin) behavior of a polariton system in 1D and 2D
quantum systems has already been investigated theoretically in a
number of works (see, for instance, Ref.~\onlinecite{Gippius}) and
was reported experimentally \cite{Baas,Paraiso,TimNPhot}. Most of
the theoretical approaches are based on solution of the GPe.
Unfortunately, this technique does
not allow to account for the dissipation dynamics of polaritons due
to interaction with the crystal lattice (phonon-mediated processes).
The density-matrix approach, which is being developed in current
manuscript, does. In the limiting case of zero temperature we
immediately reproduce the results obtained by the GPe, as expected.

\begin{figure}
\label{TempFig}
\includegraphics[width=0.95\linewidth]{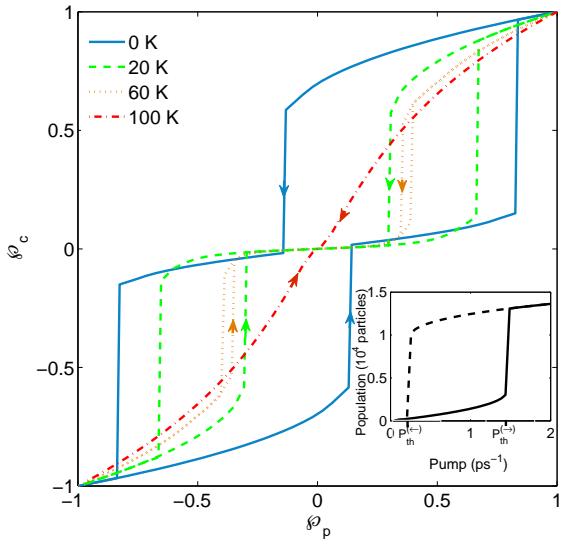}
\caption{Dependence of the circular polarization degree of the
driven polariton mode on the circular polarization degree of the
driving pump in the absence of TE-TM splitting for different
temperatures. Hysteresis loops, the signatures of the
multistability behavior, shrink with the increase of temperature
(T=0, 20, 60 K) and finally disappear at T$\approx$100 K. Inset:
dependence of polariton population versus pumping intensity for
single component system demonstrating the phenomenon of bistability
(T=0 K).} \label{fig2}
\end{figure}

The multistability (with multi-hysteresis) characteristic is shown
in Fig.~2 for different temperatures in range 1-100 K, in the
absence of TE-TM splitting. In this case, since there is no
mechanism of the transition between $\sigma^+$ and $\sigma^-$ modes,
this effect can be understood in terms of independent bistable
dynamics of $\sigma^+$ and $\sigma^-$  modes.  Accordingly, in
inset we present the population hysteresis curve of a spinless
polaritons to clarify the forthcoming discussion. Let's begin with inset. In the certain
range of pumps the polariton population can take two different
values depending on the history of the pumping process. If we slowly
increase the intensity of pump, at some threshold value
$P_{th}^{(\rightarrow)}$ the population of the ground state jumps up
abruptly due to the resonance of the blue shifted polariton energy
with the energy of the laser mode. The system keeps staying at this
high-populated state with further increase of the pump intensity. In
the backward direction, when we decrease the intensity of pump, the
bistable transition to the low-populated state appears at the lower
pump intensity ($P_{th}^{(\leftarrow)}<P_{th}^{(\rightarrow)}$) and
hysteresis curve appears.

With account for spin, polariton-polariton interaction becomes
polarization-dependent, which leads to multistability of the
polariton circular polarization (see Ref. \onlinecite{Gippius} for
the detailed discussion of the situation at $T=0$). This effect is
illustrated in main plot of Fig.~2 for different temperatures, where the pump
intensity is fixed and its circular polarization degree $\wp_p$ is
being changed. Let us explain this phenomenon with the help of above
discussion  for spinless case. Keeping the total pump intensity,
lets change its circular polarization from $\sigma^-$ ($\phi=\pi$,
and $\wp_p=-1$) to $\sigma^+$ ($\phi=0$, and $\wp_p=1$), see
Fig.~\ref{fig1}).
Initially
there exists only $\sigma^-$ polaritons
in the ground state, thus $\wp_c=-1$.  As $\wp_p$ is increased,
$\sigma^+$ component starts to become more populated, and at certain
threshold value of $\wp_p$ the first bistable jump up in $\wp_c$
occurs that implies the abrupt increase of the $\sigma^+$ component.
Further increase of $\wp_p$ leads to the second jump up of the
polarization degree $\wp_c$ due to the bistability jump down of
$\sigma^-$ component from high population state to low population
state. Finally the system reaches the state with only $\sigma^+$
component and $\wp_c=1$.
In the backward direction (decrease pump
polarization degree from +1 to -1) the first jump down is due to the
abrupt increase of $\sigma^-$ component, while the second jump down
is explained by the abrupt decrease of $\sigma^+$ component
occupancy.

With increasing temperature, the multistability loops start to
shrink and become totally destroyed at about $T\approx 100$ K. It
occurs due to the dissipation processes coming from interaction with
acoustic phonons. At higher temperature the spin-independent
polariton-phonon interaction makes the dependence $\wp_c(\wp_p)$
quasi-linear, as it should be expected, indeed, in the case when
coherent nonlinearities play no role and there is no transition
between circular polarized components the polarization degree of the
system should coincide with those of the pump.
\begin{figure}
\includegraphics[width=0.95\linewidth]{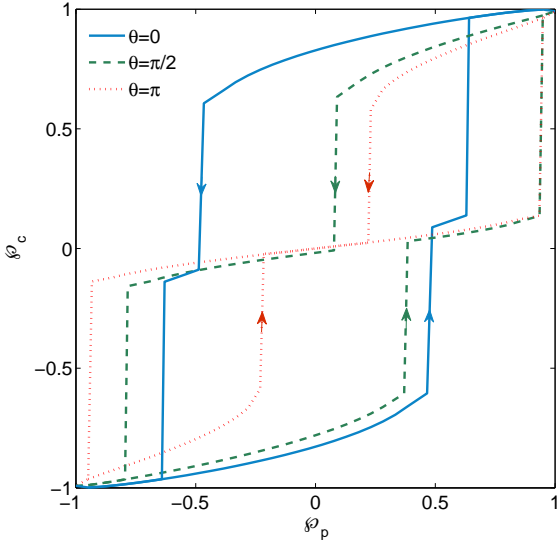}
\caption{Internal versus external circular polarization degree for
different xy-plane projections of the pseudospin of the pump
(different azimuthal angles $\theta$) for $\Omega=0.08$ meV, and T=0
K. Due to the finite value of the TE-TM coupling, the equivalency
between x and y linear polarizations is broken. Therefore, the
choice of meridian line of Poincar\'{e} sphere along which the pump
laser evolves from $\sigma^-$  to $\sigma^+$ state becomes crucial
for the polarization dynamics of the polariton system.} \label{fig3}
\end{figure}

Now let us introduce the TE-TM splitting to see its effect on
polarization multistability. The corresponding term removes the
isotropy in the xy plane since it acts as an effective magnetic
field in +x direction. Consequently, the population of each
component $n_{\bd k}^{\pm}$ becomes dependent not only on circular
polarization of the pump, but also on its in-plane component as can
be seen from Eq.~\eqref{pseudo1}. In Fig.~3, the dependence of the
internal circular polarization degree of the system $\wp_c$ is
plotted as a function of the circular polarization degree of the pump
$\wp_p$ for three different in-plane angles $\theta$ between the in-
plane pseudospin of the pump and direction along the wire: axis Ox (see
Fig.~\ref{fig1}).  Azimuthal angle $\theta$ comes in the pumping
Hamiltonian as the relative phase factor between the pumping amplitudes,
i.~e. $p_+=e^{i\theta}p_-$. It is observed that relative phase
drastically modifies the profile of the $\wp_c(\wp_p)$ plot. First,
we note that a finite y-component of the pump pseudospin ($\theta
\neq 0,\pi$) destroys the  symmetry of the multistability curve
with respect to the change of the sign of circular polarization of
the pump (see dotted green line). Also, comparing the results for
$\theta=0$ (+x direction) and $\theta=\pi$ (-x direction) cases,
we see two quite different polarization behaviors. This difference can be
understood from the first term of the kinetic equation
\eqref{pseudo2} for $\bd S_\bot$, where the two cases ($\theta=0$
and $\theta=\pi$) give contributions with opposite signs.
Therefore, the internal circular polarization degree becomes highly
sensitive to the choice of meridian on the surface of the
Poincar\'{e} sphere along which the circular polarization of the
laser is changed between $\sigma^+$ and $\sigma^-$. It should be noted,
that for circularly polarized pump the effect of rotation of pseudospin
due to TE-TM splitting is invisible: for the used value $\Omega=0.08$ meV and circular polarized pump the effect of macroscopic self-trapping plays major role \cite{SelfTrap1,SelfTrap2,SelfTrap3,ShelykhJosephson} and rotation of pseudospin is blocked. For reduced values of circular polarization self-trapping regime is lost and effects of TE-TM splitting become visible.
\begin{figure}
\includegraphics[width=0.95\linewidth]{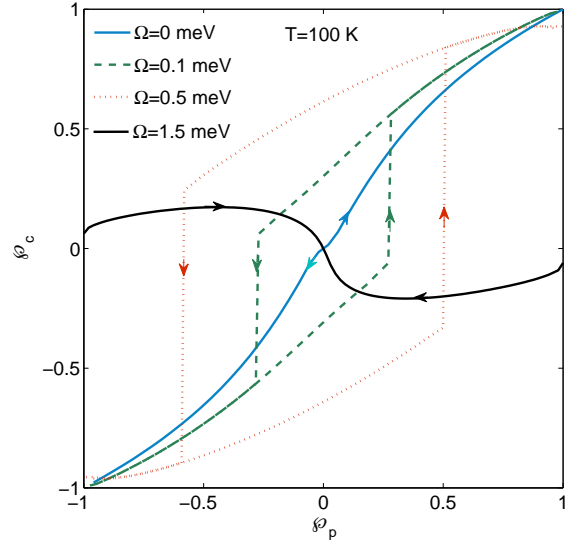}
\caption{Circular polarization degree of the
driven mode versus the circular polarization degree of pump for
different values of the longitudinal-transverse splitting $\Omega$
for T=100 K. Thus TE-TM splitting revives the bistability behavior even at
high temperature values. At strong TE-TM splittings, i. e., $\Omega=1.5$ meV, the circular polarizability
diminishes due to strong mixing of $\sigma^\pm$ modes, and the hysteresis loop disappears.} \label{fig4}
\end{figure}

Let us finally analyze the combined effect of the TE-TM splitting and
scattering on phonons (in the rest of the calculations the in-plane component
of the pump pseudospin is taken along the +x direction, i.e., $\theta=0$).
As it was shown before in Fig.~2, in the absence of
TE-TM splitting due to the dissipative nature of polariton-phonon
interactions the hysteresis behavior is washed out at 100 K.
On the other hand, if $\Omega\neq 0$, as in Fig.~4, the bistable behavior
can be recovered and bistability phenomena can survive up to
higher temperatures as compare to the case $\Omega=0$.
However, instead of two-stepped hysteresis loop shown in Fig.~\ref{fig3}, we
observe only one-stepped behavior. This result suggests the transition
from two-independent modes dynamics (the two modes are the mode
with $\sigma^+$ and the mode with $\sigma^-$ polarizations) to a single collective mode
dynamics. In fact, at some critical value of the TE-TM splitting,
(around 0.1 meV in our parameter regime), transition from
high population state to low population state of the one mode is always
accompanied by the simultaneous transition from low population state
to high population state of the other mode and crossover from the multistable behavior to the bistable occurs.

If one increases the value of the TE-TM splitting field even further, polaritons would prefer to stay in quasi-linearly polarized state due to strong mixing of $\sigma^\pm$ modes, even for pumping by the fully circularly polarized laser. This situation occurs at $\Omega=1.5$ meV in Fig.~4, where polaritons become highly linearly polarized even at the values of $\wp_p=\pm1$ due to high value of the effective magnetic field in the +x direction. The last term in Eq.~\eqref{pseudo2} is responsible for this behavior. The particles align their pseudospin parallel to the strong effective magnetic field to minimize the total energy in the system. Meanwhile, the hysteresis behavior vanishes, and the difference between the backward and forward swappings disappears.


\section{Conclusions}
In conclusion, we developed a formalism for the description of the
dissipative dynamics of an inhomogeneous spinor polariton system in
real space and time accounting for polariton-polariton interactions,
polariton-phonon scattering and effect of the TE-TM effective
magnetic field. We applied our formalism to one-dimensional
polariton condensate at different temperatures to investigate the
dynamics of the circular polarization of the system when it is
driven by the external homogeneous laser pump. We showed that the
polarization of the condensate is highly sensitive not only to the
history of the strength of the pump, but also to the phase of the
elliptical polarization degree of this pump. In the presence of TE-TM
field we observe the survival of this phenomena up to very high
temperatures.

\section{Acknowledgements}
We thank E. B. Magnusson for useful discussions and help with
numerical calculations. The work was supported by Rannis "Center of
excellence in polaritonics" and FP7 IRSES project "SPINMET". I.G.S. and I.A.S. thank International Institute of Physics for hospitality.


\end{document}